\documentclass[preprint,showpacs,showkeys,preprintnumbers,amsmath,amssymb,floatfix,nofootinbib,prc]{revtex4-1}

\usepackage{color}   
\usepackage{graphicx}
\usepackage{dcolumn}

\begin{document}

\def\bb    #1{\hbox{\boldmath${#1}$}}

\title{Multiplicity fluctuations in the Glauber Monte Carlo approach}

\author{Maksym Deliyergiyev}
\email{maksym.deliyergiyev@ujk.edu.pl}
\affiliation{Institute of Physics, Jan Kochanowski University, 25-406 Kielce, Poland}

\author{Maciej Rybczy\'nski}
\email{maciej.rybczynski@ujk.edu.pl}
\affiliation{Institute of Physics, Jan Kochanowski University, 25-406 Kielce, Poland}

\begin{abstract}
We discuss multiplicity fluctuations of charged particles produced in nuclear collisions measured event-by-event by the NA49 experiment at CERN SPS within the Glauber Monte Carlo approach. We use the concepts of wounded nucleons and wounded quarks in the mechanism of multiparticle production to characterize multiplicity fluctuations expressed by the scaled variance of multiplicity distribution. Although Wounded Nucleon Model correctly reproduce the centrality dependence of the average multiplicity in Pb+Pb collisions, it completely fails in description of corresponding centrality dependence of scaled variance of multiplicity distribution. Using subnucleonic degrees of freedom, i.e. wounded quarks within Wounded Quark Model, it is possible to describe quite well the multiplicity distribution of charged particles produced in proton+proton interactions. However, the Wounded Quark Model with parameters describing multiplicity distribution of particles produced in proton+proton interactions substantially exceeds the average multiplicity of charged particles produced in Pb+Pb collisions. To obtain values of average multiplicities close to those experimentally measured in Pb+Pb collisions, the concept of shadowed quark sources is implemented. Wounded Quark Model with implemented shadowing source scenario reproduces the centrality dependence of scaled variance of multiplicity distribution of charged particles produced in Pb+Pb collisions in the range from the most central to mid-peripheral interactions. 
\end{abstract}
\keywords{fluctuations, multiplicity, multiparticle production, Glauber Monte Carlo}
\pacs{02.50.Ey, 05.10.Ln, 12.40.Ee}

\maketitle

\section{Introduction}
\label{sec:intro}

Particle interactions in collisions of relativistic ions usually lead to production of secondary particles whose number rises with increasing collision energy. The multiplicity of produced charged particles, $N$ is one of the fundamental observables being an important characteristic of the global properties of the system.

Fluctuations of particle multiplicity, mean transverse momentum ($\langle p_{T} \rangle$), transverse energy ($E_{T}$) and other global observables in heavy ion collisions have become in nowadays one of the most important topics of interest, since they provide some relevant signals for the formation of quark-gluon plasma (QGP). Using a substantial number of particles produced in collisions of relativistic ions in the Super Proton Synchrotron (SPS) and the Large Hadron Collider (LHC) at the European Organization for Nuclear Research (CERN) and the Relativistic Heavy Ion Collider (RHIC) at Brookhaven National Laboratory (BNL), one can investigate fluctuations of various observables which may be sensitive to the transitions between hadronic and partonic phases~\cite{Stephanov:1999zu,Koch:2008ia} with the use the event-by-event method~\cite{Heiselberg:2000fk}.

Fluctuations in the initial conditions are essential to the full understanding of the dynamics of collisions of relativistic ions. The simplest way of modeling these fluctuations is based on random selection of positions of nucleons in each nucleus before the collision and deterministic designation, obtained after a collision, energy density according to the made assumptions. That is what happens, for example, in the standard Glauber Monte Carlo (GMC) model~\cite{Miller:2007ri} or Kharzeev-Levin-Nardi (KLN) Monte Carlo~\cite{Kharzeev:2002ei}, which have been used for many years. 

Recently, the PHENIX Collaboration suggested~\cite{Adler:2013aqf} that the Wounded Quark Model (WQM)~\cite{Eremin:2003qn,Bialas:1977en,Anisovich:1977av} works better than the popular Wounded Nucleon Model (WNM)~\cite{Bialas:1976ed,Bialas:2008zza}, in particular in description of average multiplicities. In Ref.~\cite{Eremin:2003qn} it was shown that data on multiplicities obtained at RHIC may be reproduced within a WQM. Larger number of constituents and a decreased quark-quark cross section with respect to nucleons, allow to obtain an approximately linearly increased particle production at midrapidity as a function of wounded quarks number, $dN/d\eta\propto Q_{W}$. The quark scaling for the SPS energies was discussed in Ref.~\cite{KumarNetrakanti:2004ym}. The agreement with the data may be achieved without the introduction of the binary-collision component~\cite{Kharzeev:2000ph,Back:2001xy}, which introduces non-linearity between the number of nucleons participating in collision and the multiplicity of produced particles. 

In this paper we are interested to examine an impact of subnucleonic components of matter on the dynamics of the early stage of the collision of relativistic nuclei. We  extend and generalize the approach proposed in~\cite{Bozek:2016kpf} to describe the centrality dependence of multiplicity fluctuations of charged particles registered by the NA49 experiment at CERN SPS~\cite{Alt:2006jr}. Quite unexpectedly, the measured scaled variance of the multiplicity distribution in Pb+Pb collisions show quite non-trivial dependence on collision centrality~\cite{Alt:2006jr}. It is close to unity at very central collisions, however it shows a substantial difference from unity at peripheral interactions. This effect is not present in a commonly used models of nuclear collisions. In Ref.~\cite{Alt:2006jr} the NA49 data on centrality dependence of scaled variance of the multiplicity distribution was compared to HIJING~\cite{Gyulassy:1994ew}, HSD~\cite{Cassing:1999es}, UrQMD~\cite{Bleicher:1999xi}, and VENUS~\cite{Werner:1993uh} simulations. A detailed discussion of HSD and UrQMD predictions for centrality dependence of scaled variance was also presented in~\cite{Konchakovski:2005hq}. The models produce approximately Poissonian~\footnote{Variance (second central moment) of Poisson distribution (PD) equals its mean value thus scaled variance of PD equals to unity.} multiplicity distributions independent of centrality. Although there are some models trying to describe the non-monotonic behavior of the scaled variance of multiplicity distribution as a function of collision centrality~\cite{Rybczynski:2004zi,Cunqueiro:2005hx,Gazdzicki:2005rr} up to now there is no commonly accepted explanation of this phenomenon.

The article is organized as follows. In Sec.~\ref{sec:exp_data} we discuss the experimental data used in this paper, in Sec.~\ref{sec:brief} we briefly depict the idea of GMC approach of description of relativistic nuclear collisions. The resultant multiplicity fluctuations obtained in WNM are discussed in Sec.~\ref{sec:mult_WNM}. The next section is devoted to analysis of the multiplicity fluctuations within WQM. Finally, Sec.~\ref{sec:cr} contains our summary, with the conclusion that WQM works much better in description of charged particles multiplicity fluctuations than WNM. In the Appendix we use a simple example do demonstrate the influence of fluctuating number of sources for the final multiplicity distribution.

\section{Multiplicity fluctuations data}
\label{sec:exp_data}

In this paper the charged particle multiplicity distribution $P\left(N\right)$ and its scaled variance $\omega$ is used to describe multiplicity fluctuations. Here $P\left(N\right)$ denote the probability to detect a charged particle multiplicity $N$ in the single event of high energy nuclear collision. $P\left(N\right)$ is normalized to unity by definition, $\sum_{N} P\left(N\right)=1$. The scaled variance of multiplicity distribution (the so-called Fano-factor), $\omega\left(N\right)$ provides a suitable index for the degree of deviation from a Poisson distribution~\footnote{If $\omega > 1$, the distribution is called to be over-dispersed or super-Poissonian, namely the existence of clusters of occurrences may happen; if $\omega < 1 $, it is said the distribution to be under-dispersed or sub-Poissonian, namely this situation relates to arrangements of occurrences that are more ordinary than the randomness connected with a Poisson process. For the Poisson distribution, $\omega=1$.} and is defined as:
\begin{equation}
\omega\left(N\right)=\frac{Var\left(N\right)}{\langle N\rangle}=\frac{\langle N^{2}\rangle - \langle N\rangle^{2}}{\langle N\rangle},
\label{eq:omega_def}
\end{equation}
where $Var\left(N\right)=\sum_{N}\left(N-\langle N\rangle\right)^{2}\cdot P\left(N\right)$ is the variance of the distribution and $\langle N\rangle=\sum_{N} N\cdot P\left(N\right)$ is the average multiplicity.

In some models of nuclear collisions the scaled variance of multiplicity distribution do not depend on the number of sources of particle production. Commonly used models of nuclear interactions, the models of superposition, are constructed using the concept of particle production by the independent sources. In these models the scaled variance has two contributions. The first is due to the fluctuations of the number of particles emitted by a single source $\omega_{s}$, the second is due to the fluctuations in the number of sources $\omega_{k}$:
\begin{equation}
\omega=\omega_{s}+ \langle N_{s}\rangle \omega_{k},
\label{eq:ScaledVariance_TwoContr}
\end{equation}
where $\langle N_{s}\rangle$ is the mean multiplicity of hadrons from a single source. The participant nucleons of a collision are considered to be proportional to the sources of particle production. In order to minimize the fluctuations of the number of sources, the centrality variation in the ensemble of events should be as small as possible. However, it is \textit{a~priori} not known how the fluctuations of the number of projectile and target nucleons participating in collision contribute to the multiplicity fluctuations in different regions of the phase-space. There are several theoretical concepts which drive the multiplicity fluctuations such as resonance decays, fluctuations in relativistic gases, string-hadronic models, onset of deconfinement and critical point. 

In the often used superposition model which is the WNM~\cite{Bialas:1976ed}, the sources are wounded nucleons, i.e. the nucleons that have to collide at least once (calculated usually with use of GMC approach). In WNM, the scaled variance in nucleus-nucleus collisions have the same value as in nucleon-nucleon interactions under condition of fixed number of wounded nucleons.

The NA49 experiment located at CERN SPS published data on system-size and centrality dependence of fluctuations of the number of charged particles produced in proton+proton and Pb+Pb collisions at $\sqrt{s_{NN}}=17.3$~GeV~\cite{Alt:2006jr}. The NA49 detector registered multiplicity distributions of particles produced in the restricted rapidity interval $1.1 < y_{\pi} < 2.6$~\footnote{$y_{\pi}$ means rapidity calculated under assumption of mass of $\pi$ meson.} in the centre of mass frame. The azimuthal acceptance of the NA49 detector was also limited. Such restrictions correspond to a fraction of about 17\% of accepted charged particles~\cite{Alt:2006jr}. The particles produced in both proton+proton and Pb+Pb collisions were measured at exactly the same experimental acceptance. This makes a unique possibility to describe both proton+proton and Pb+Pb data in exactly the same way without introduction any additional biases.

The NA49 is a fixed-target experiment and is equipped in the forward calorimeter allowing for precise determination of the number of nucleons-spectators~\footnote{In the present work by the \textit{nucleons-spectators} we call those nucleons which did not interact with other nucleons during the collision, and \textit{nucleons-participants} are those nucleons which suffer at least \textit{one} inelastic collision.}, $N_{s}^{proj}$ from projectile nucleus. Thus, the number of projectile nucleons participating in the collision, $N_{p}^{proj}$ may be calculated as $N_{p}^{proj}=A-N_{s}^{proj}$, where $A$ is the atomic mass number of projectile nucleus. In the NA49 experiment $N_{p}^{proj}$ was used as a measure of centrality in the nucleus-nucleus collisions. To avoid unnecessary contribution from fluctuations of $N_{p}^{proj}$ to the observed multiplicity distributions the resultant multiplicity distributions of charged articles were obtained at fixed number of $N_{p}^{proj}$. 

However, in~\cite{Gazdzicki:2005rr} it was observed that even with a fixed number of participants from projectile nucleus, the number of target participants fluctuates and it was suggested that the observed sizable multiplicity fluctuations in the forward rapidity domain of Pb+Pb collisions are due to fluctuations of the number of participants from the target. In~\cite{Gazdzicki:2005rr} it was assumed that the nucleons participants in the target (projectile) nucleus contribute to the projectile (target) nucleus fragmentation region, which means that these areas overlap. The authors of Ref.~\cite{Gazdzicki:2005rr} called it a {\it mixing} model, in contrast to the {\it transparency} model in which the projectile participants contribute only to the projectile fragmentation region, and the target participants into the target fragmentation region. The transparency model is compatible with the limiting fragmentation hypothesis~\footnote{Hypothesis of limiting fragmentation states that for a sufficiently high collision energy particle production becomes target and energy independent in the projectile (target) fragmentation domain corresponding to the rapidities close to that of the projectile (target).}~\cite{Benecke:1969sh} while the mixing model contradicts it. Both models should obviously be treated as idealizations. The analysis of d+Au collisions at RHIC~\cite{Bialas:2004su} shows that in reality we have both mixing and transparency. In~\cite{a-13} it was shown that the target nucleus does not affect the multiplicity distributions of particles produced in proton+Pb minimum bias collisions in the forward rapidity region and the effect of the increased multiplicity fluctuations due to influence of target participants is not observed. The results were compared with the corresponding results obtained in the proton+proton interactions and the predictions of models. Fig. 5 of Ref.~\cite{a-13} clearly shows that the transparency model describes the data quite well, but the mixing model seems to be excluded. Of course, the mechanism of the particle production in Pb+Pb collisions may be different than in proton+Pb interactions. Nevertheless, the results of~\cite{a-13} indicate that there is no strong mixing of the projectile and target nucleus fragmentation regions, at least in proton+nucleus collisions.

\section{Glauber-like models and Glauber Monte Carlo approach}
\label{sec:brief}

The Glauber model~\cite{Glauber:1959aa} was released almost seventy years ago to find description of high-energy collision of atomic nuclei treated as composite structures. Until then there was no systematic calculations regarding nuclear systems as projectile or target. Glauber model containing quantum theory of collisions of composite particles allows to describe experimental results on collisions of protons with deuterons and heavier nuclei. In the mid 1970s Bialas {\it et al}~\cite{Bialas:1976ed,Bialas:1977pd} used Glauber model to describe inelastic nuclear collisions in their WNM. Bialas {\it et al}~\cite{Bialas:1976ed} formulation allows to treat a collision between nuclei as a superposition of incoherent collisions between their nucleons. A review of Glauber modelling of high-energy collisions was given in~\cite{Miller:2007ri}. 

In the last years the popular GMC approach became an important tool in the analysis of collisions of relativistic ions~\cite{Miller:2007ri}. One of the most important application of the GMC simulation is the estimate of the number of participants dependence on the centrality, especially in the collider experiments~\cite{Miller:2007ri,Loizides:2014vua,Broniowski:2007nz,Rybczynski:2013yba,Bozek:2019wyr}. The presence of the event-by-event fluctuations in the initial Glauber phase is a very important aspect of the approach. These fluctuations are transferred to the distributions of the experimentally registered hadrons. GMC initial state is often used as an starting point for event-by-event hydrodynamics~\cite{Andrade:2006yh}.

The multiplicity of particles produced in nuclear collisions fluctuates event-by-event. In the GMC approach using wounded nucleons or wounded partons only a part of these fluctuations can be described by the fluctuations of the number of sources emitting particles (nucleons or partons). In order to describe the experimentally observed charged hadrons multiplicity distributions the model multiplicity distribution should be expressed as a convolution of the distribution of the number of emitting sources, $P_{S}$ with the distribution, $P_{H}$, of hadrons emitted from a singe source. So, the number of charged particles is given by:
\begin{equation}
N=\sum_{i=1}^{N_{p}} n_{i},
\label{eq:n_def}
\end{equation}
where $n_{i}$ follows from $P_{H}$ with the generating function $H\left(z\right)$ and $N_p$ comes from $P_{S}$ with the generating function $S\left(z\right)$. Thus, the measured multiplicity distribution $P\left(N\right)$ is given by the compound generating function
\begin{equation}
G\left(z\right)=S\left[H\left(z\right)\right]
\label{eq:comp_gen}
\end{equation}
and finally
\begin{equation}
P\left(N\right)=\frac{1}{N!}\frac{d^{N}G\left(z\right)}{dz^{N}}\Biggl\vert_{z=0}.
\label{eq:pn_def}
\end{equation}

To parameterize $P_H$ we use Negative Binomial (NB) distribution which is a statistical tool frequently used to describe multiplicity distributions of particles produced in nuclear collisions:
\begin{equation}
P_{NB}\left(N,\langle N\rangle,k\right)=\binom{N+k-1}{N}\left(\frac{\langle N\rangle}{k}\right)^{N}\left(1+\frac{\langle N\rangle}{k}\right)^{-N-k}.
\label{eq:nbd_def}
\end{equation}
NB has two free parameters: $\langle N\rangle$ describing mean multiplicity and, not necessarily integer parameter $k$ ($k\geq 1$) affecting shape of the distribution. 

In this article we use GLISSANDO~\cite{Broniowski:2007nz,Rybczynski:2013yba,Bozek:2019wyr} which is a versatile GMC generator for early-stages of relativistic ion collisions, including the wounded nucleon and wounded quark (in general wounded parton) models, with possible admixture of binary collisions. A state of the art inelastic nucleon-nucleon collision profile is implemented. A statistical distribution of the strength of the sources can be overlaid on the distribution of sources. For the purposes defined in this work we implement the shadowing procedure as described in~\cite{Chatterjee:2015aja} into the recent version of GLISSANDO~\cite{Bozek:2019wyr}.

\section{Multiplicity fluctuations in Wounded Nucleon Model}
\label{sec:mult_WNM}

\begin{figure}
\begin{center}
\includegraphics[width=0.49 \textwidth]{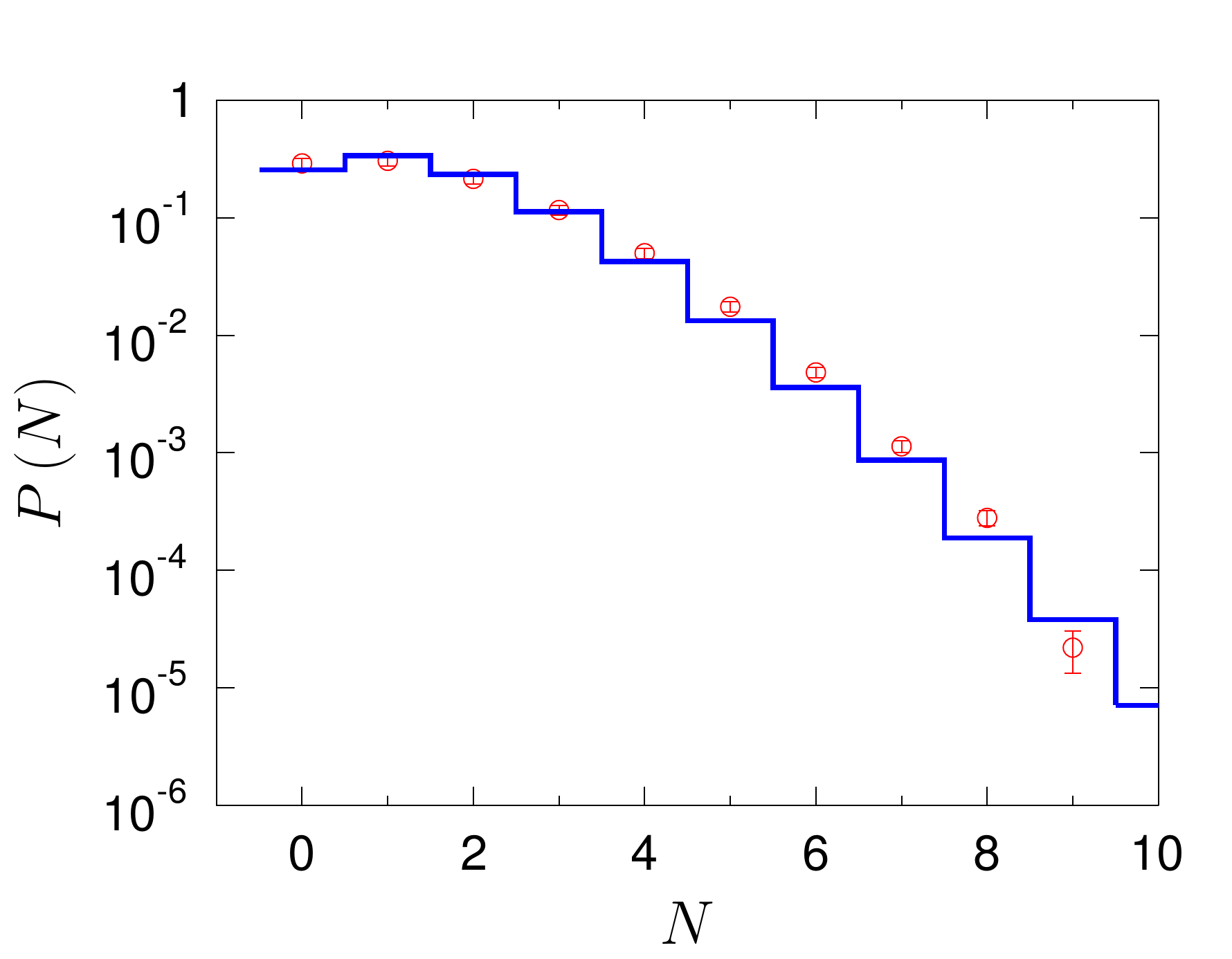}

\end{center}
\vspace{-5mm}
\caption{(Color online) Multiplicity distribution of charged hadrons produced in proton+proton interaction and registered by the NA49 experiment~\cite{Alt:2006jr} (circles). Histogram shows the Negative Binomial distribution fit with parameters $\langle N_{NB}\rangle=1.4$ and $k=9.8$.} 
\label{fig:negbin_pp}
\end{figure} 
\begin{figure*}
\begin{center}
\includegraphics[width=0.49 \textwidth]{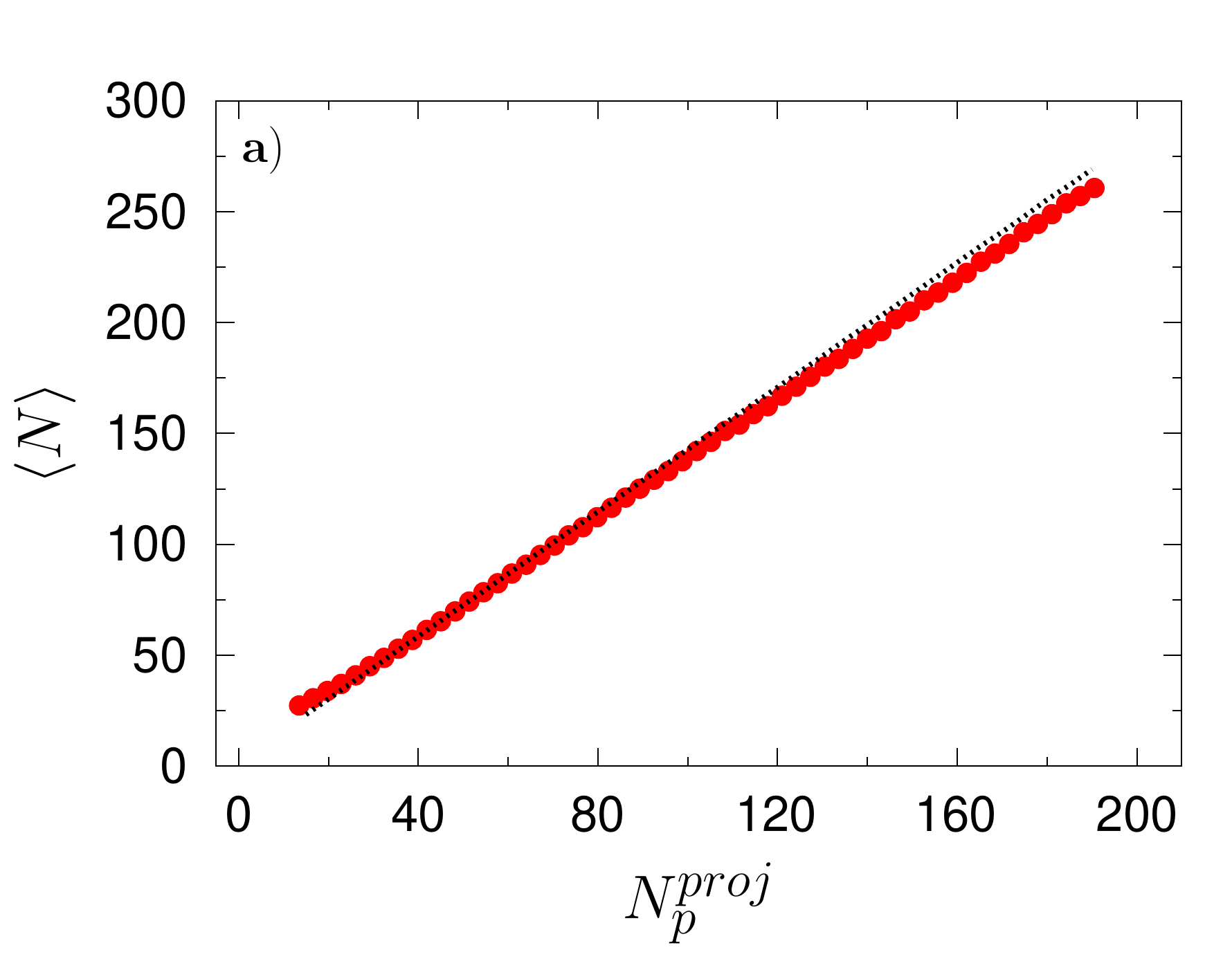} 
\includegraphics[width=0.49 \textwidth]{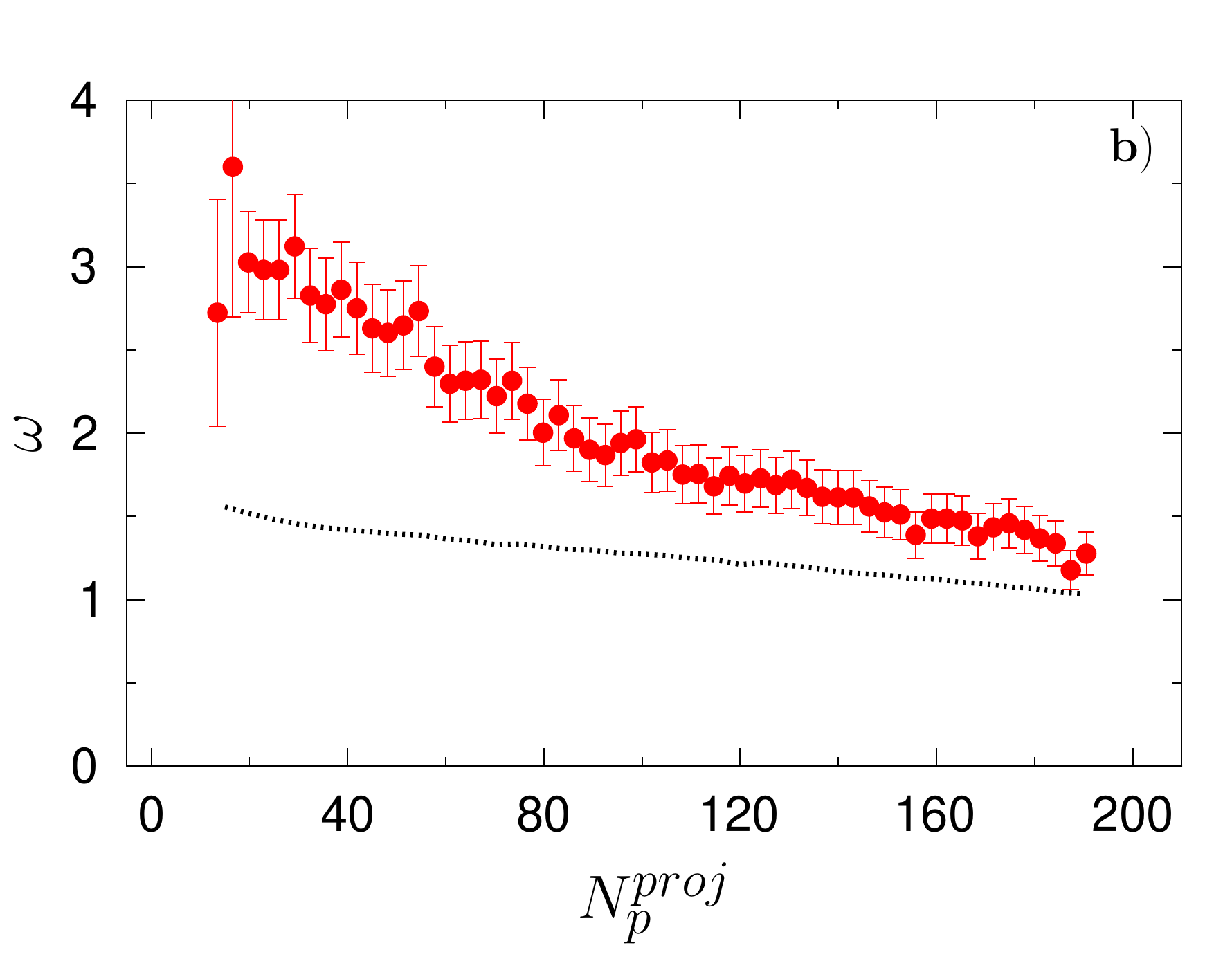} 
\end{center}
\vspace{-5mm}
\caption{(Color online) Average number of charged hadrons produced in Pb+Pb collisions (panel a)) and corresponding scaled variance (panel b)) of charged hadrons multiplicity distribution registered by the NA49 experiment~\cite{Alt:2006jr} plotted as a function of number of nucleons from the projectile nucleus participating in the collision (circles). Dotted lines show the results from WNM.} 
\label{fig:res_wnm}
\end{figure*} 

The NA49 data on charged hadrons multiplicity distributions in proton+proton and centrality selected Pb+Pb collisions were obtained using the same experimental conditions. The basic concept of description these results within WNM is then as follows. First, we fit the experimental proton+proton multiplicity distribution by the NB distribution. Then, we overlay the distribution of emitting sources, which are wounded nucleons with the NB distribution with the parameters obtained from the fit to proton+proton data. The NB fit to proton+proton data is based on the $\chi^2$ method with the errors taken from~\cite{Alt:2006jr,a-13}. The value of $\chi^2$ divided by the number of degrees of freedom, $\chi^2/N_{\rm{dof}}=3.3$, indicates the rather good quality of the fit. NB fit roughly describes experimental proton+proton multiplicity distribution. It is illustrated in the Fig.~\ref{fig:negbin_pp}. The fit provides us the following NB parameters, $\langle N_{NB}\rangle=1.4$ and $k=9.8$. This corresponds to the variance of the NB fit, $Var\left(N_{NB}\right)=1.6$.

To reproduce the experimental multiplicity distributions in centrality selected Pb+Pb collisions, we prepared multiplicity distributions at the fixed number of wounded nucleons from projectile nucleus, similarly as in the NA49 experiment. Fig.~\ref{fig:res_wnm} shows the results from the model compared to the data. WNM quite well reproduces the average charged multiplicity at all centralities, what is shown in the panel a). However, WNM completely fails in reproduction of the corresponding centrality dependence of the scaled variance of multiplicity distribution, see panel b) of Fig.~\ref{fig:res_wnm}. A small monotonic increase of the WNM scaled variance with decreasing number of nucleons participating in the collision, with respect to the experimental data may be explained by the small contribution from the fluctuations of target participants, whose number cannot be fixed experimentally.

\section{Multiplicity fluctuations in Wounded Quark Model}
\label{sec:mult_WQM}

\begin{figure}
\begin{center}
\includegraphics[width=0.49 \textwidth]{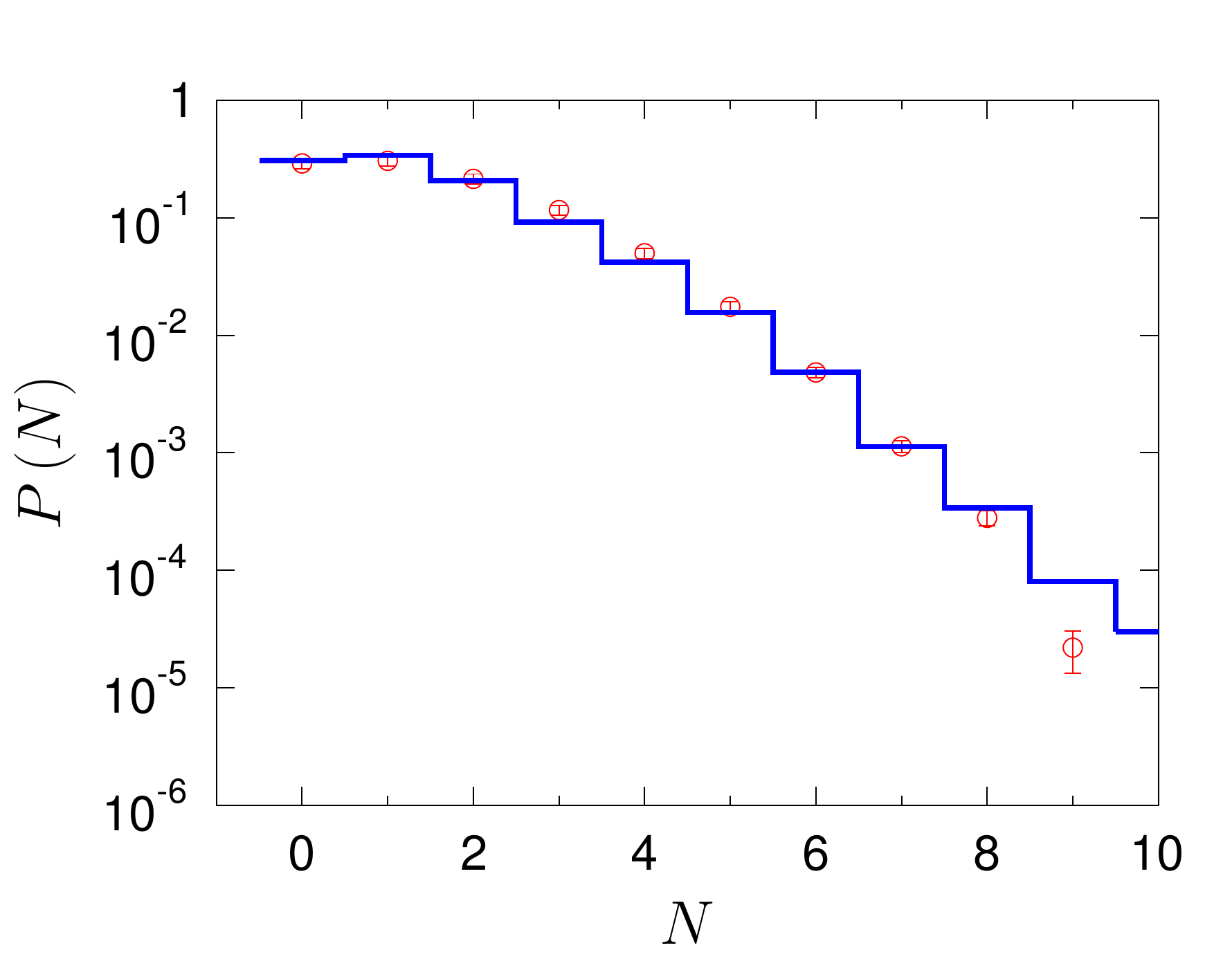} 
\end{center}
\vspace{-5mm}
\caption{(Color online) Multiplicity distribution of charged hadrons produced in proton+proton interactions and registered by the NA49 experiment~\cite{Alt:2006jr} (circles). Histogram shows the WQM fit. See text for details.} 
\label{fig:wqm_pp}
\end{figure}

Using subnucleonic degrees of freedom it is possible to make an analysis of the proton+proton interactions. The proton+proton inelastic collision profile as well as total inelastic cross section in the WQM is described using quark-quark collisions~\cite{Bozek:2019wyr,Bozek:2016kpf}. The average number of wounded quarks per nucleon in proton+proton interactions at the considered energy is $Q_{W}=1.27$. The charged particles multiplicity distribution is a result of a convolution of the distribution of particles produced by each wounded quark and the distribution of the number of wounded quarks. In the Fig.~\ref{fig:wqm_pp} we present the resultant fit of the NA49 proton+proton multiplicity distribution with the WQM predictions. To obtain it we again use the NB distribution given by Eq.~\eqref{eq:nbd_def} as $P_{H}$, now with the parameters $\langle N_{NB}\rangle=0.53$ and $k=14$. The quality of the fit is rather poor. However, within this description, the differences between observed and expected values are acceptable, $\chi^2/N_{\rm{dof}}\simeq 6$.

\begin{figure}
\begin{center}
\includegraphics[width=0.49 \textwidth]{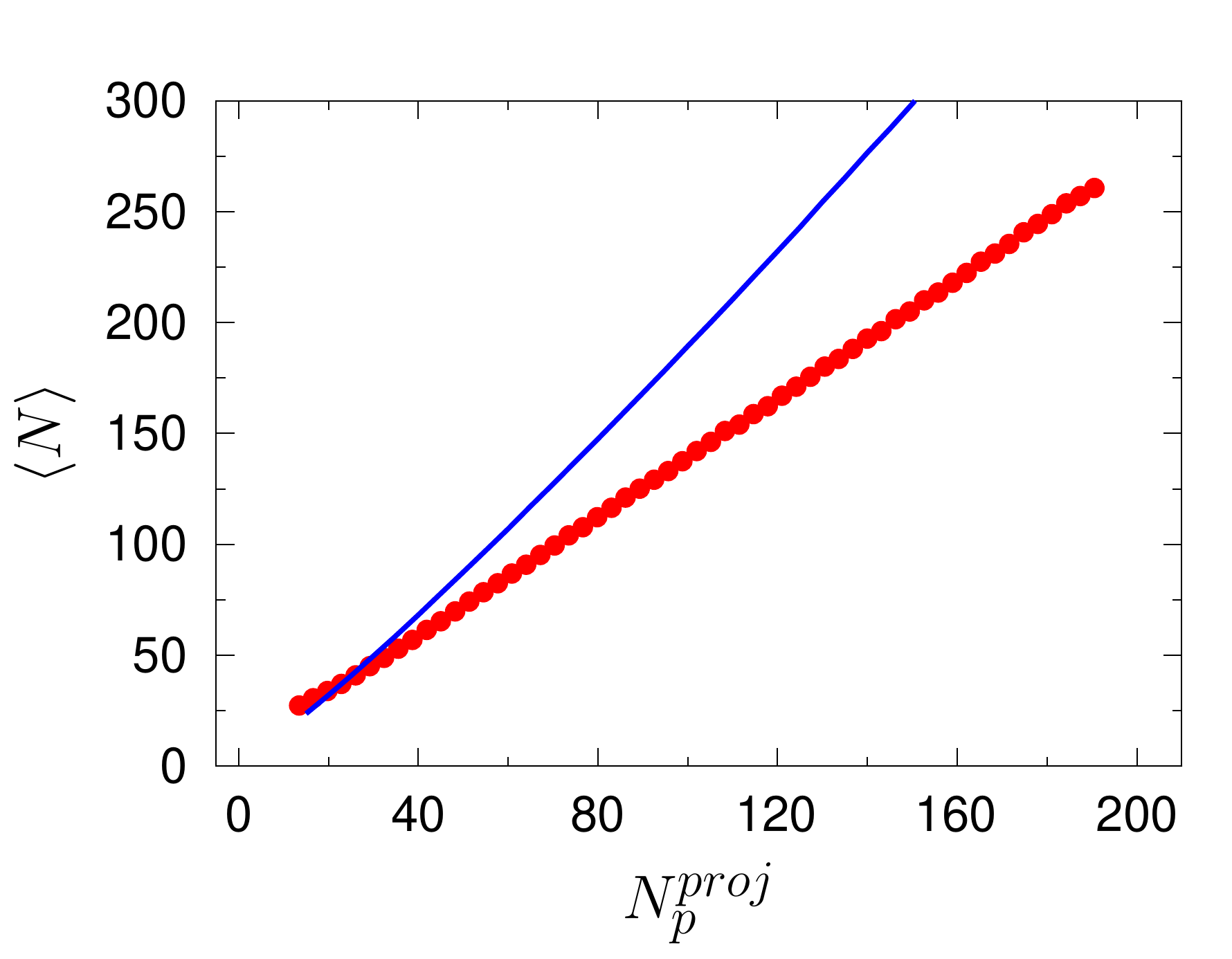} 

\end{center}
\vspace{-5mm}
\caption{(Color online) Average number of charged hadrons produced in Pb+Pb collisions registered by the NA49 experiment~\cite{Alt:2006jr} plotted as a function of number of projectile participants (circles). Line shows the results from the WQM.} 
\label{fig:wqm_av_mult}
\end{figure} 

\begin{figure}
\begin{center}
\includegraphics[width=0.49 \textwidth]{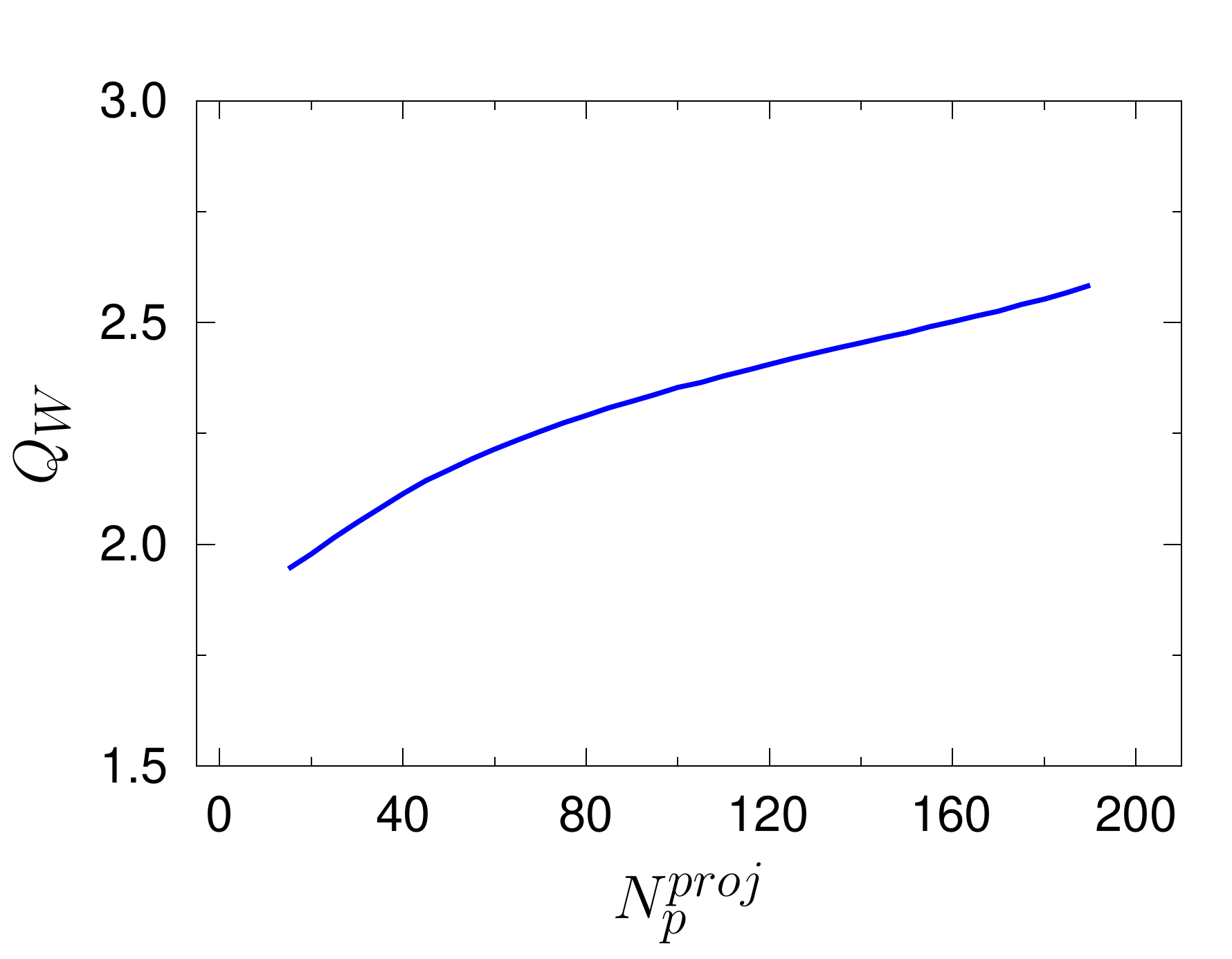} 
\end{center}
\vspace{-5mm}
\caption{(Color online) Average number of wounded quarks per nucleon as a function of number of projectile participants in Pb+Pb collisions.} 
\label{fig:wqm_av_qq}
\end{figure}

Similarly as in the case of WNM we prepared centrality selected Pb+Pb collisions using NB parameters obtained from the WQM fit to proton+proton data. Unfortunately, WQM cannot reproduce properly the centrality dependence of the average multiplicity, see Fig.~\ref{fig:wqm_av_mult}. This is caused by the higher average number of wounded quarks per nucleon in centrality selected Pb+Pb collisions in comparison to proton+proton interactions, see Fig.~\ref{fig:wqm_av_qq}. The mean value of wounded quarks is highest in the most central Pb+Pb collisions and decreases slowly when going to peripheral. However it is always higher than the corresponding number in proton+proton interactions.

\begin{figure}
\begin{center}
\includegraphics[width=0.49 \textwidth]{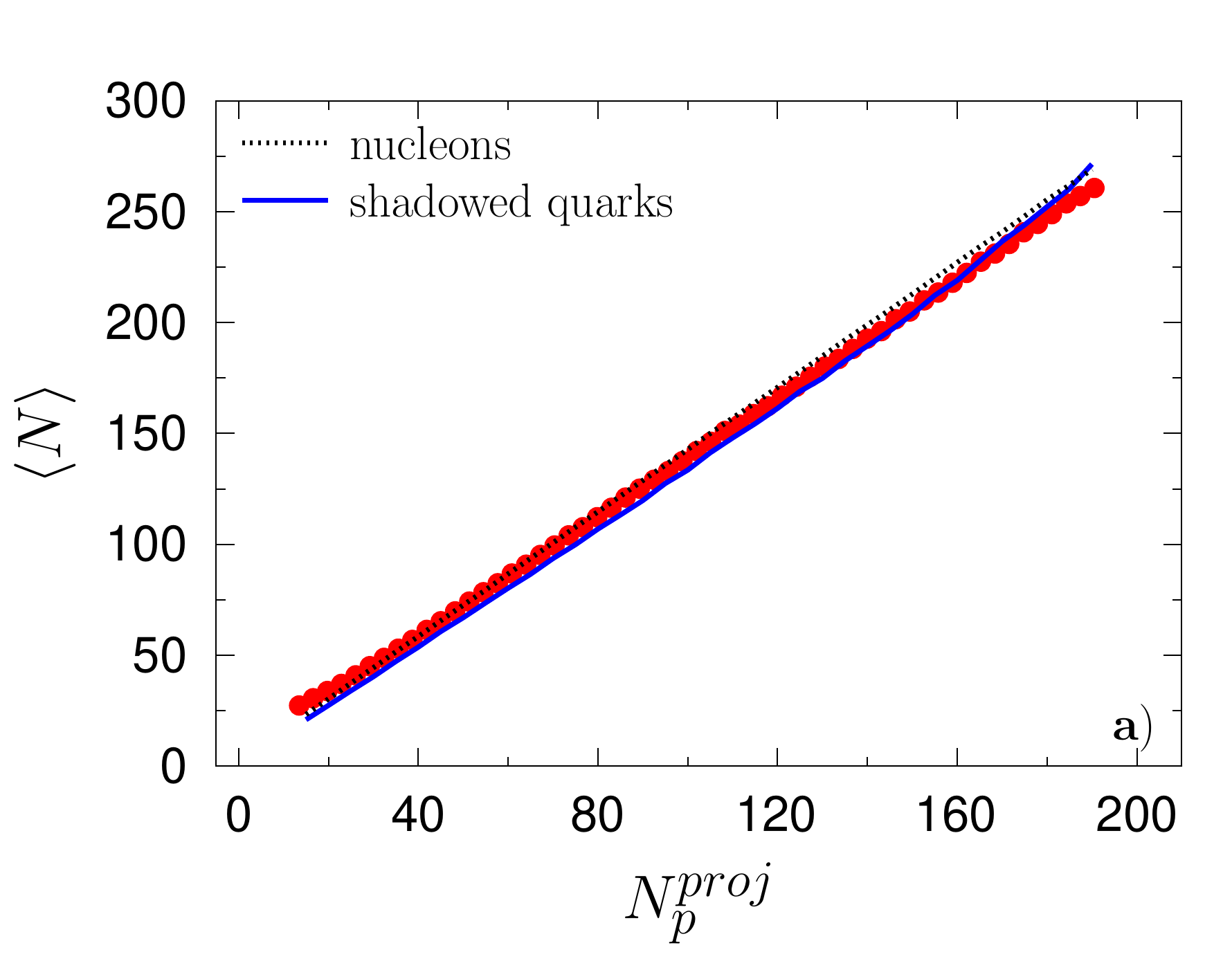} 
\includegraphics[width=0.49 \textwidth]{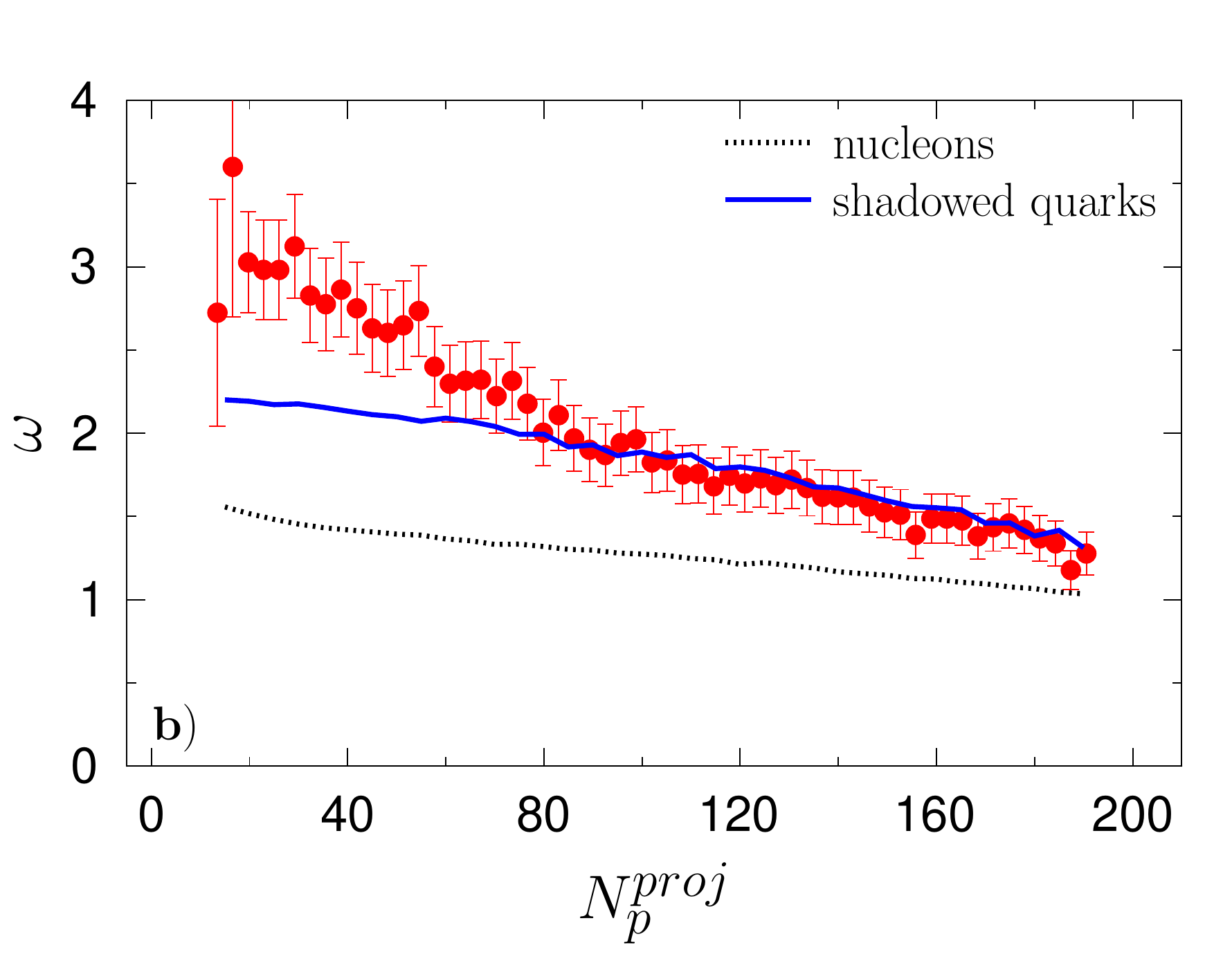} 
\end{center}
\vspace{-5mm}
\caption{(Color online) Average number of charged hadrons produced in Pb+Pb collisions (panel a)) and corresponding scaled variance (panel b)) of charged hadrons multiplicity distribution registered by the NA49 experiment~\cite{Alt:2006jr} plotted as a function of number of projectile participants (circles). Full lines show the results of shadowed WQM. With dotted lines we indicate the results from WNM.} 
\label{fig:wqm_shadow_res}
\end{figure} 

In order to decrease the average charged multiplicity we implement the source shadowing procedure as presented by Chatterjee {\it et al} in~\cite{Chatterjee:2015aja}. In short: the contribution to particle production by quark sources located inside the nucleus is shadowed by those being in front. The particle production suppression factor $S\left(n,\lambda\right)$ for the contribution from a quark source shadowed by the $n$ other quark sources from the same nucleus ahead is
\begin{equation}
S\left(n,\lambda\right)=\exp\left(-n\cdot\lambda\right)
\label{eq:supp}
\end{equation}
with $\lambda$ being a phenomenological parameter. Such idea of shadowing was discussed first in Ref.~\cite{Glauber:1955qq}. In the left panel of Fig.~\ref{fig:wqm_shadow_res} we demonstrate the resultant centrality dependence of the average multiplicity in Pb+Pb collisions after implementation of the shadowing with the $\lambda=0.95$. Right panel of the Fig.~\ref{fig:wqm_shadow_res} shows the corresponding centrality dependence of the scaled variance of charged multiplicity distribution. We show results from WQM with shadowed quarks together with the results from WNM. We note substantial increase of the value of scaled variance of multiplicity distributions for collisions at all centralities when shadowed quarks are used in comparison to the standard WNM predictions. Such increase of fluctuations is caused by the presence of additional source of fluctuations which is the fluctuating number of wounded quarks in nucleon, see Appendix for the discussion.

There are also possible other variants of shadowing. Namely, instead of decreased production from the shadowed source as discussed above, such source may emit particles with a certain probability, proportional to suppression factor, given by Eq.~\eqref{eq:supp}. We also checked such mechanism but we did not find differences in comparison to the discussed above shadowing scenario.


\section{Concluding remarks}
\label{sec:cr}

The main purpose of the study presented in this article is to check the influence of nucleonic and subnucleonic degrees of freedom for the dynamics of the early stage of the collision of relativistic nuclei. We use the concept of wounded nucleons and wounded quarks in the mechanism of multiparticle production to describe charged particles multiplicity fluctuations expressed by the scaled variance of multiplicity distribution and observed in collisions of relativistic ions by the NA49 experiment at CERN SPS. We take the opportunity that NA49 data on multiplicity fluctuations for proton+proton and centrality selected Pb+Pb collisions were obtained at the same experimental acceptance. Wounded nuleons and wounded quarks are implemented using Glauber Monte Carlo approach. Our results are as follows.
\begin{itemize}

\item WNM describe reasonably well the centrality dependence of the average multiplicity of charged particles produced in Pb+Pb collisions.
It is possible when distribution of wounded nucleons is overlaid with the NB distribution with parameters describing multiplicity distribution in proton+proton interactions.

\item WNM does not describe the centrality dependence of the scaled variance of multiplicity distribution of charged particles produced in Pb+Pb collisions at any centrality.

\item Using subnucleonic degrees of freedom, i.e. wounded quarks within WQM, it is possible to describe quite well the multiplicity distribution of charged particles produced in proton+proton interactions.

\item However, the WQM with parameters describing multiplicity distribution of particles produced in proton+proton interactions substantially exceeds the average multiplicity of charged particles produced in Pb+Pb collisions. This is due to higher average number of wounded quarks per nucleon in centrality selected Pb+Pb collisions with respect to proton+proton interactions.

\item To obtain values of average multiplicities close to those experimentally measured in Pb+Pb collisions, the concept of shadowed quark sources was implemented. In this scenario each quark source which is behind other source in the same nucleus emit less number of particles, proportionally to the number of shadowing sources.

\item WQM with implemented shadowing source scenario partially reproduces the centrality dependence of scaled variance of multiplicity distribution of charged particles produced in Pb+Pb collisions. The WQM predictions reasonably agree with the data for the central and mid-peripheral Pb+Pb collisions, with number of projectile participants in the range $80<N_p^{proj}<200$. For more peripheral collisions the discrepancy between data and WQM predictions grows when going towards peripheral collisions. The substantial increase of the value of scaled variance of multiplicity distribution in the WQM with respect to WNM predictions is caused by the additional fluctuations, namely the fluctuations of the number of quark sources in the nucleon. 

\item WQM shows that implementation of sources, like quarks, helps to describe heavy ion data in the range $80<N_p^{proj}<200$. This leads to idea that there should be a triggering process that cause an additional increase in number of sources (sea quarks/gluons) in the range of $0<N_p^{proj}<80$, which is smoothly switching off beyond this threshold.


\end{itemize}


\vspace*{0.3cm}
\centerline{\bf Acknowledgements}
\vspace*{0.3cm}
The numerical simulations were carried out in laboratories created under the project
``Development of research base of specialized laboratories of public universities in Swietokrzyskie region'',
POIG 02.2.00-26-023/08, 19 May 2009.\\
This research  was supported by the National Science Centre (NCN) grant 2016/23/B/ST2/00692.

\appendix*

\section{Multiplicity fluctuations in compound distributions}
\label{app:fluct_wq}

In this simple example we show the influence of the fluctuating number of sources for the resultant multiplicity distribution.
Let $P_{1}\left(n\right)$ be the multiplicity distribution of particles emitted by the single source. We compare results from two scenarios:
\begin{enumerate}
\item Final multiplicity distribution is the sum of constant number of $P_{1}\left(N\right)$ distributions,
\item Final multiplicity distribution is the sum of fluctuating number of $P_{1}\left(N\right)$ distributions.
\end{enumerate}

The multiplicity $N$ of generated particles is given by Eq.~\eqref{eq:n_def}. The average multiplicity is then:
\begin{equation}
\langle N\rangle=\langle N_{p}\rangle\langle n\rangle,
\label{eq:mean}
\end{equation}
where $\langle n\rangle$ is the average multiplicity from single source. Variance of multiplicity distribution:
\begin{equation}
Var\left(N\right)=\langle N_{p}\rangle Var\left(n\right)+Var\left(N_{p}\right)\langle n\rangle^{2},
\label{eq:variance}
\end{equation}
where $Var\left(n\right)$ means variance of the multiplicity distribution from a single source and $Var\left(N_{p}\right)$ is the variance of the number of sources distribution.

\begin{figure}
\begin{center}
\includegraphics[width=0.49 \textwidth]{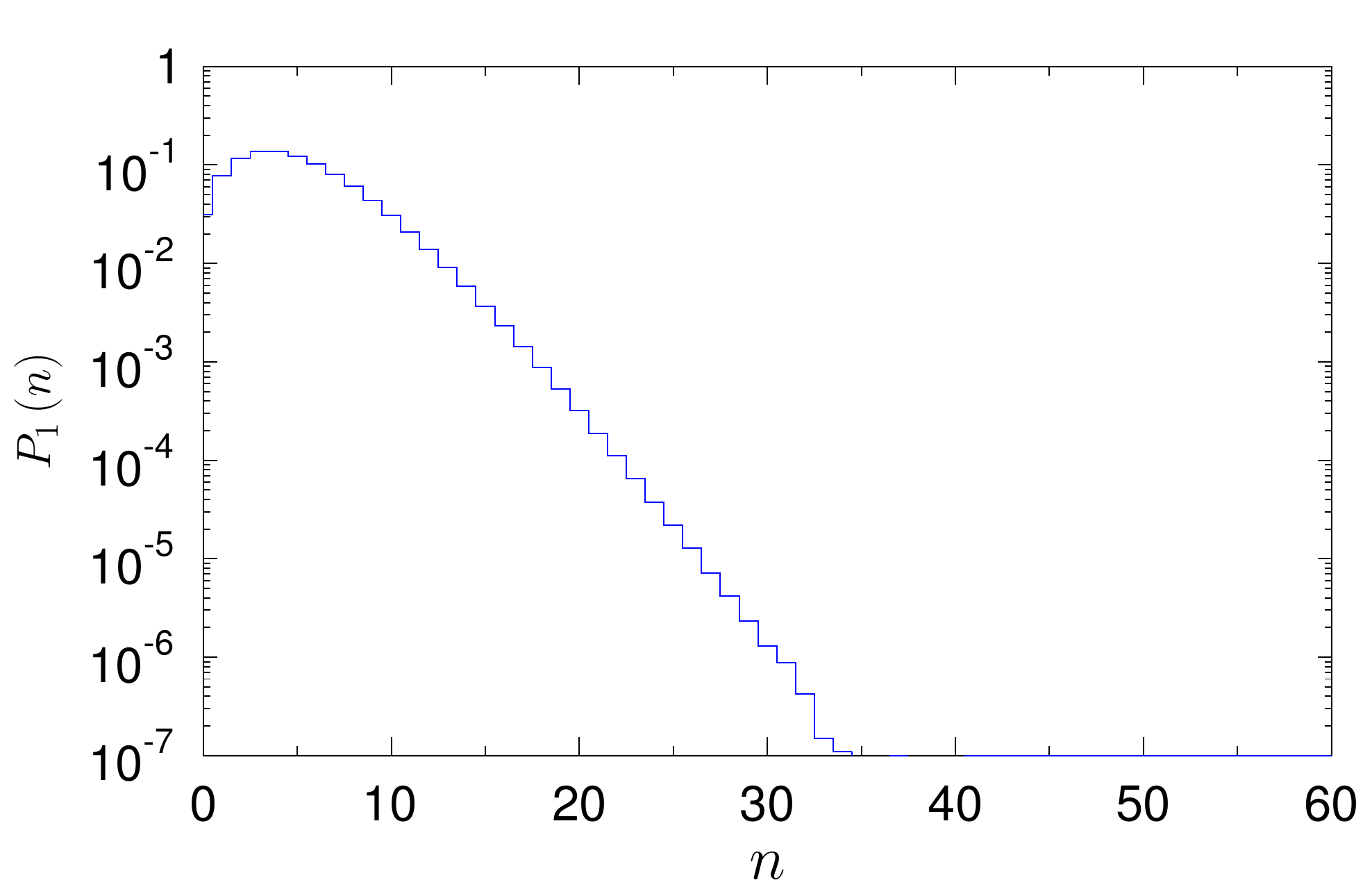} 
\end{center}
\vspace{-5mm}
\caption{(Color online) Multiplicity distribution of particles emitted from a single source, given by NB distribution with the average $\langle n\rangle=5$ and shape parameter $k=5$ what corresponds to variance, $Var\left(n\right)=10$.} 
\label{fig:nbd_single}
\end{figure} 

\begin{figure}
\begin{center}
\includegraphics[width=0.49 \textwidth]{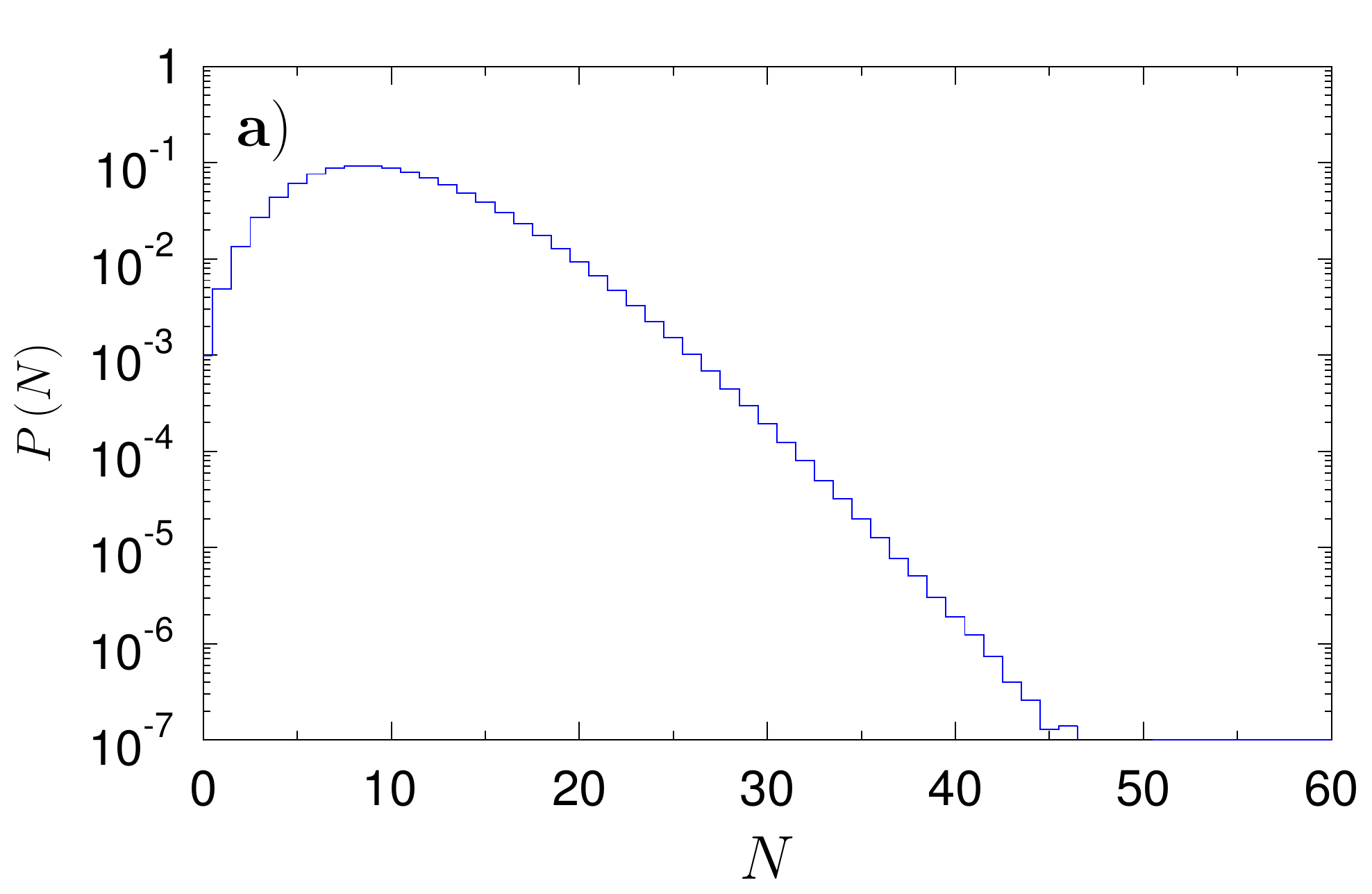} 
\includegraphics[width=0.49 \textwidth]{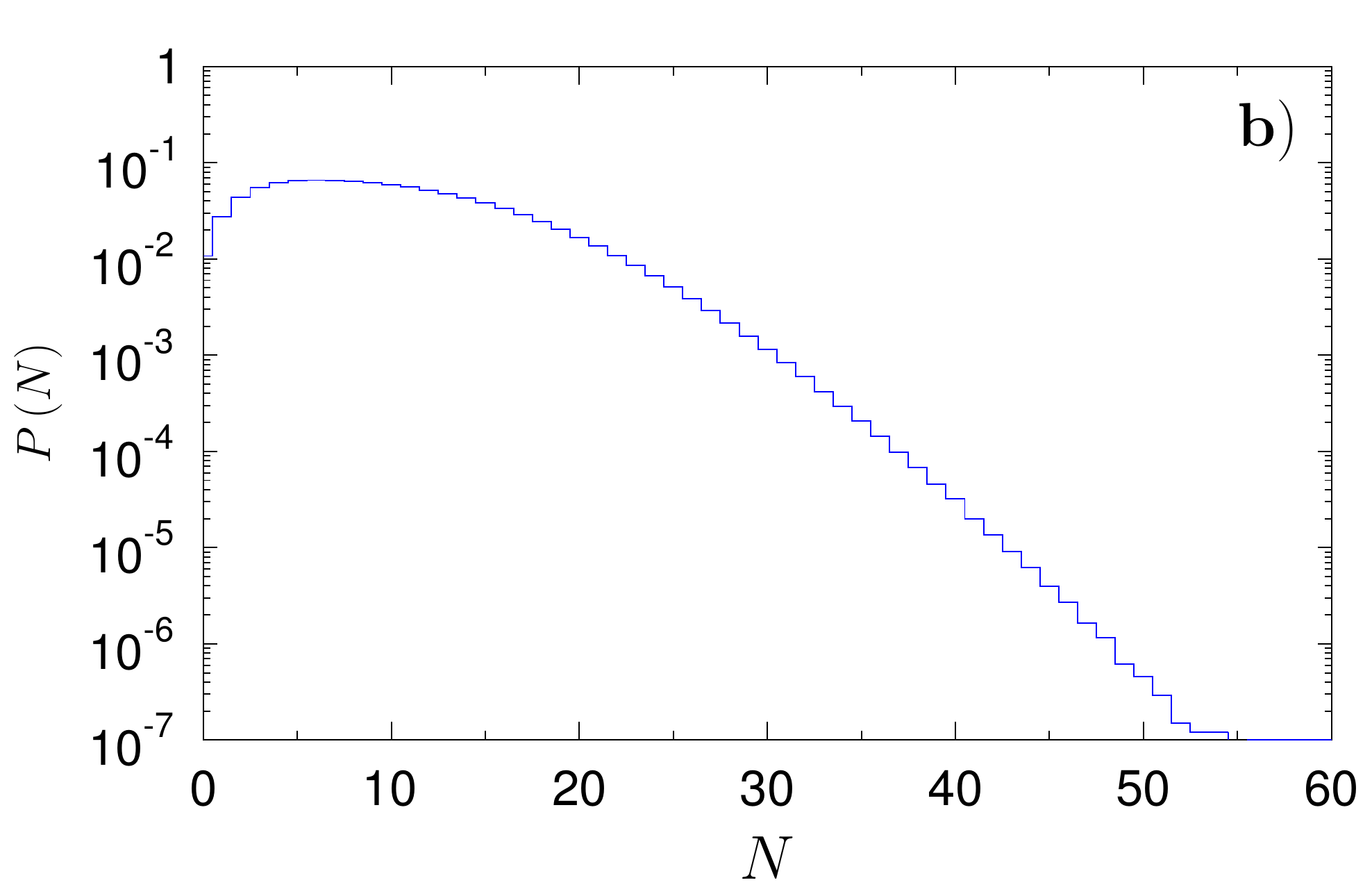} 
\end{center}
\vspace{-5mm}
\caption{(Color online) Multiplicity distributions of particles emitted from: a constant number $N_{p}=2$ sources (panel a)) and fluctuating number of sources with $\langle N_{p}\rangle=2$ and $Var\left(N_{p}\right)=0.667$. In both cases distribution from the single source is the same, given by NB distribution with the average $\langle n\rangle=5$ and shape parameter $k=5$. See text for details.} 
\label{fig:nbd_comp}
\end{figure} 

In the first scenario we generate the sum of the $N_{p}=2$ numbers generated from $P_{1}\left(n\right)$ distribution. In the second scenario we sum the $N_{p}$ numbers given by $P_{1}\left(n\right)$, but $N_{p}$ is generated from the set of $N_{p}\in\{1,2,3\}$ uniformly generated numbers. Obviously $\langle N_{p}\rangle=2$ in both scenarios. Also average multiplicities $\langle N\rangle$ have the same values in both scenarios. However variances of the resultant multiplicity distributions are substantially different since in the first scenario $Var\left(N_{p}\right)=0$ and in the second one $Var\left(N_{p}\right)=0.667$. 

As an example we provide results from Monte Carlo simulations of the two discussed scenarios. $P_{1}\left(n\right)$ distribution is given by NB distribution with  average value $\langle n\rangle=5$ and variance $Var\left(n\right)=10$, what corresponds to shape parameter $k=5$, see Fig.~\ref{fig:nbd_single}. The final multiplicity distribution in the first scenario have $\langle N\rangle=10$ and $Var\left(N\right)=20$, and the corresponding shape parameter $k=10$. In the second scenario the resultant distribution have the same mean value, $\langle N\rangle=10$ but is much broader than distribution from first scenario, its variance $Var\left(N\right)=36.68$ ($k=3.75$) what is expected for fluctuating number of sources, Eq.~\eqref{eq:variance}.


\begin{thebibliography}{99}

\bibitem{Stephanov:1999zu} 
M.~A.~Stephanov, K.~Rajagopal, E.~V.~Shuryak,
Phys.\ Rev. {\bf D60}, 114028 (1999)
doi:10.1103/PhysRevD.60.114028
[hep-ph/9903292].

\bibitem{Koch:2008ia} 
V.~Koch,
Relativistic Heavy Ion Physics,
Chapter of the book "Relativistic Heavy Ion Physics,
Stock (Ed.), Springer, Heidelberg, 2010, p. 626-652.
[nucl-th/0810.2520].

\bibitem{Heiselberg:2000fk} 
  H.~Heiselberg,
  Phys.\ Rept.\  {\bf 351}, 161 (2001)
  doi:10.1016/S0370-1573(00)00140-X
  [nucl-th/0003046].


\bibitem{Miller:2007ri} 
  M.~L.~Miller, K.~Reygers, S.~J.~Sanders and P.~Steinberg,
  Ann.\ Rev.\ Nucl.\ Part.\ Sci.\  {\bf 57}, 205 (2007)
  [nucl-ex/0701025].
  
\bibitem{Kharzeev:2002ei} 
  D.~Kharzeev, E.~Levin and M.~Nardi,
  Nucl.\ Phys.\ A {\bf 730}, 448 (2004)
  Erratum: [Nucl.\ Phys.\ A {\bf 743}, 329 (2004)]
  doi:10.1016/j.nuclphysa.2004.06.022, 10.1016/j.nuclphysa.2003.08.031
  [hep-ph/0212316].

\bibitem{Adler:2013aqf} 
  S.~S.~Adler {\it et al.} [PHENIX Collaboration],
  Phys.\ Rev.\ C {\bf 89}, no. 4, 044905 (2014)
  doi:10.1103/PhysRevC.89.044905
  [arXiv:1312.6676 [nucl-ex]].

\bibitem{Eremin:2003qn} 
  S.~Eremin and S.~Voloshin,
  Phys.\ Rev.\ C {\bf 67}, 064905 (2003)
  doi:10.1103/PhysRevC.67.064905
  [nucl-th/0302071].

\bibitem{Bialas:1977en} 
  A.~Bialas, W.~Czyz and W.~Furmanski,
  Acta Phys.\ Polon.\ B {\bf 8}, 585 (1977).

\bibitem{Anisovich:1977av} 
  V.~V.~Anisovich, Y.~M.~Shabelski and V.~M.~Shekhter,
  Nucl.\ Phys.\ B {\bf 133}, 477 (1978).
  doi:10.1016/0550-3213(78)90237-7

\bibitem{Bialas:1976ed} 
  A.~Bialas, M.~Bleszynski and W.~Czyz,
  Nucl.\ Phys.\ B {\bf 111}, 461 (1976).
  doi:10.1016/0550-3213(76)90329-1

\bibitem{Bialas:2008zza} 
  A.~Bialas,
  J.\ Phys.\ G {\bf 35}, 044053 (2008).
  doi:10.1088/0954-3899/35/4/044053

\bibitem{KumarNetrakanti:2004ym} 
  P.~Kumar Netrakanti and B.~Mohanty,
  Phys.\ Rev.\ C {\bf 70}, 027901 (2004)
  doi:10.1103/PhysRevC.70.027901
  [nucl-ex/0401036].

\bibitem{Kharzeev:2000ph} 
  D.~Kharzeev and M.~Nardi,
  Phys.\ Lett.\ B {\bf 507}, 121 (2001)
  doi:10.1016/S0370-2693(01)00457-9
  [nucl-th/0012025].

\bibitem{Back:2001xy} 
  B.~B.~Back {\it et al.} [PHOBOS Collaboration],
  Phys.\ Rev.\ C {\bf 65}, 031901 (2002)
  doi:10.1103/PhysRevC.65.031901
  [nucl-ex/0105011].

\bibitem{Bozek:2016kpf} 
  P.~Bożek, W.~Broniowski and M.~Rybczyński,
  Phys.\ Rev.\ C {\bf 94}, no. 1, 014902 (2016)
  doi:10.1103/PhysRevC.94.014902
  [arXiv:1604.07697 [nucl-th]].

\bibitem{Alt:2006jr} 
  C.~Alt {\it et al.} [NA49 Collaboration],
  Phys.\ Rev.\ C {\bf 75}, 064904 (2007)
  doi:10.1103/PhysRevC.75.064904
  [nucl-ex/0612010].

\bibitem{Gyulassy:1994ew} 
  M.~Gyulassy and X.~N.~Wang,
  Comput.\ Phys.\ Commun.\  {\bf 83}, 307 (1994)
  doi:10.1016/0010-4655(94)90057-4
  [nucl-th/9502021].

\bibitem{Back:2004zf} 
B.B.~Back,
J.\ Phys.\ Conf.\ Ser. {\bf 5}, 1-16 (2005)
doi:10.1088/1742-6596/5/1/001
[nucl-ex/0411012].

\bibitem{Cassing:1999es} 
  W.~Cassing and E.~L.~Bratkovskaya,
  Phys.\ Rept.\  {\bf 308}, 65 (1999).
  doi:10.1016/S0370-1573(98)00028-3

\bibitem{Bleicher:1999xi} 
  M.~Bleicher {\it et al.},
  J.\ Phys.\ G {\bf 25}, 1859 (1999)
  doi:10.1088/0954-3899/25/9/308
  [hep-ph/9909407].

\bibitem{Werner:1993uh} 
  K.~Werner,
  Phys.\ Rept.\  {\bf 232}, 87 (1993).
  doi:10.1016/0370-1573(93)90078-R

\bibitem{Konchakovski:2005hq} 
  V.~P.~Konchakovski, S.~Haussler, M.~I.~Gorenstein, E.~L.~Bratkovskaya, M.~Bleicher and H.~Stoecker,
  Phys.\ Rev.\ C {\bf 73}, 034902 (2006)
  doi:10.1103/PhysRevC.73.034902
  [nucl-th/0511083].

\bibitem{Rybczynski:2004zi} 
  M.~Rybczynski and Z.~Wlodarczyk,
  J.\ Phys.\ Conf.\ Ser.\  {\bf 5}, 238 (2005)
  doi:10.1088/1742-6596/5/1/022
  [nucl-th/0408023].

\bibitem{Cunqueiro:2005hx} 
  L.~Cunqueiro, E.~G.~Ferreiro, F.~del Moral and C.~Pajares,
  Phys.\ Rev.\ C {\bf 72}, 024907 (2005)
  doi:10.1103/PhysRevC.72.024907
  [hep-ph/0505197].

\bibitem{Gazdzicki:2005rr} 
  M.~Gazdzicki and M.~I.~Gorenstein,
  Phys.\ Lett.\ B {\bf 640}, 155 (2006)
  doi:10.1016/j.physletb.2006.07.044
  [hep-ph/0511058].

\bibitem{Benecke:1969sh} 
  J.~Benecke, T.~T.~Chou, C.~N.~Yang and E.~Yen,
  Phys.\ Rev.\  {\bf 188}, 2159 (1969).
  doi:10.1103/PhysRev.188.2159

\bibitem{Bialas:2004su}
A.~Bialas and W.~Czyz,
Acta Phys.\ Polon.\  B {\bf 36}, 905 (2005)
[arXiv:hep-ph/0410265].

\bibitem{a-13}
M. Rybczynski (for the NA49 Collaboration){~in 
{\it Proceedings of the 43. International Symposium on Multiparticle Dynamics} 
S.~Chekanov and Z.~Sullivan eds., (IIT Press, Chicago, 2013) p.265, ISBN 978-1-61597-002-5}.

\bibitem{Glauber:1959aa}
  R.J.~Glauber{~in {\it Lectures in Theoretical Physics} W.~E. Brittin and
  L.~G. Dunham eds., (Interscience, New York, 1959) Vol. 1, p. 315}. 

\bibitem{Bialas:1977pd} 
  A.~Bialas, M.~Bleszynski and W.~Czyz,
  Acta Phys.\ Polon.\ B {\bf 8}, 389 (1977).

\bibitem{Loizides:2014vua} 
  C.~Loizides, J.~Nagle and P.~Steinberg,
  SoftwareX {\bf 1-2}, 13 (2015)
  doi:10.1016/j.softx.2015.05.001
  [arXiv:1408.2549 [nucl-ex]].

\bibitem{Broniowski:2007nz} 
  W.~Broniowski, M.~Rybczynski and P.~Bozek,
  Comput.\ Phys.\ Commun.\  {\bf 180}, 69 (2009)
  doi:10.1016/j.cpc.2008.07.016
  [arXiv:0710.5731 [nucl-th]].

\bibitem{Rybczynski:2013yba} 
  M.~Rybczynski, G.~Stefanek, W.~Broniowski and P.~Bozek,
  Comput.\ Phys.\ Commun.\  {\bf 185}, 1759 (2014)
  doi:10.1016/j.cpc.2014.02.016
  [arXiv:1310.5475 [nucl-th]].

\bibitem{Bozek:2019wyr} 
  P.~Bozek, W.~Broniowski, M.~Rybczynski and G.~Stefanek,
  arXiv:1901.04484 [nucl-th].

\bibitem{Andrade:2006yh} 
  R.~Andrade, F.~Grassi, Y.~Hama, T.~Kodama and O.~Socolowski, Jr.,
  Phys.\ Rev.\ Lett.\  {\bf 97}, 202302 (2006)
  doi:10.1103/PhysRevLett.97.202302
  [nucl-th/0608067].

\bibitem{Chatterjee:2015aja} 
  S.~Chatterjee, S.~K.~Singh, S.~Ghosh, M.~Hasanujjaman, J.~Alam and S.~Sarkar,
  Phys.\ Lett.\ B {\bf 758}, 269 (2016)
  doi:10.1016/j.physletb.2016.05.022
  [arXiv:1510.01311 [nucl-th]].

\bibitem{Glauber:1955qq} 
  R.~J.~Glauber,
  Phys.\ Rev.\  {\bf 100}, 242 (1955).
  doi:10.1103/PhysRev.100.242


\end{thebibliography}
\end{document}